\documentclass[
  ,final            
  ]
  {aipproc}

\layoutstyle{6x9}

\begin{document}

%
\title{Implications of Standard-Model flavor violation for new physics searches}



\author{Susan Gardner}{address={Department of Physics \& Astronomy, University of Kentucky, Lexington, KY 40506-0055 USA
}}

\begin{abstract}
I discuss far-flung ramifications of light-quark flavor physics for 
searches for physics beyond the Standard Model and consider, particularly, its role 
in the interpretation of CP-violating observables in meson decays. 
\end{abstract}

\maketitle


\section{Introduction}

The differing masses and electroweak couplings of the quarks give
rise to appreciably distinct effects 
and hence to ``flavor physics.'' 
The quarks vary widely in mass; indeed, the very large value of the top quark mass
drives the hierarchy problem 
and speaks to a resolution which also involves  
flavor. In this contribution I set aside consideration 
of the possible flavor structure of new physics
and discuss, rather, how the light-quark flavor dynamics of the Standard Model (SM) 
can impact the interpretation 
of low-energy experiments designed to identify departures from the SM. 

The decipherment of the flavor and spin structure of the proton and neutron
continues to be of intense interest, and at this meeting it has been a
central theme: many have discussed the role of strange quarks or that of the
up and down quarks, as well as of charge-symmetry breaking (CSB), 
in nucleon observables. 
The flavor physics of the light quarks 
also has broad implications for the search for physics beyond the SM. 
I will offer an overview of these and then focus on its implications for the
study of CP violation. 

\section{The Search for New Physics}

There is much theoretical ``evidence'' that the Standard Model 
is incomplete --- it leaves many questions unanswered. For example, there is no natural 
reason for the precise value of the weak mass scale, nor for the observed pattern of 
the fermions and their masses and mixing. The model 
does not incorporate gravity by design, and 
it offers no explanation for either dark matter or dark energy.  
Most notably, the Standard Model only explains some 5\% of the known 
Universe~\cite{Komatsu:2010fb}. 

\begin{figure}
  \includegraphics[height=.25\textheight]{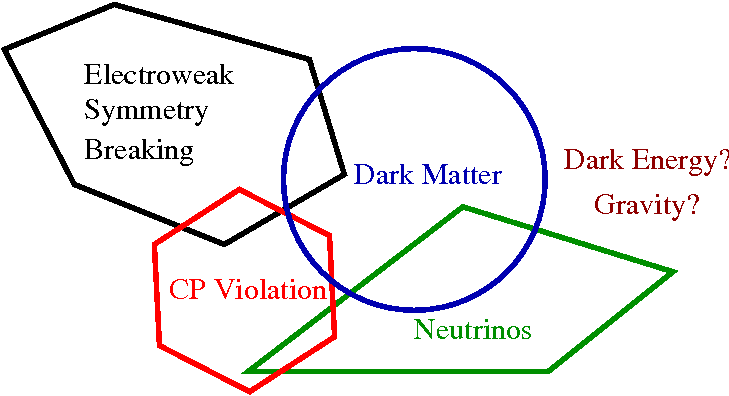}
  \caption{Themes in the search for new physics. How does flavor enter?}
\label{inter}
\end{figure}

Searches for new physics, which may well redress these 
limitations, can be organized along different themes, as shown in Fig.~\ref{inter}. 
The figure means to illustrate the possible interconnectedness of these efforts. For example, 
the origin of ``dark matter'' may ultimately be resolved~\cite{Feng:2010gw,Kaplan:1991ah}, 
all or in part, by a mechanism 
related to electroweak symmetry breaking, or by a mechanism related to that
which explains the baryon asymmetry of the universe, or by a sterile neutrino, 
or by a completely unrelated --- and perhaps as yet unknown --- mechanism.
Let us consider 
how light-quark flavor physics enters in the interpretation of 
low-energy experiments which probe these possibilities.  

\section{The Many Threads of Flavor Physics}

At this meeting we have already seen examples of how flavor physics can
impact the search for new physics. 
 In regards to the mechanism
of electroweak symmetry breaking, both Londergan~\cite{Londergan} and 
Clo{\"e}t et al.~\cite{Cloet} have studied 
the impact of CSB and medium effects, respectively, on the determination 
$\sin^2 \theta_W$ from the NuTeV experiment~\cite{nutev}. Moreover, Young has studied 
the impact of strange quarks in the proton on the 
determination of the weak couplings of the $u$ and $d$ quarks
from parity-violating electron scattering~\cite{Young}.  
In each case the flavor-breaking effects required explicit evaluation. 

The strange-quark structure of the nucleon also impacts the hunt for dark matter. In
particular, it impacts the interpretation of 
the weakly-interacting massive particle (WIMP) exclusion plots which emerge from 
direct searches for anomalous nuclear recoils from WIMP-nucleon scattering~\cite{gaitskell}. 
A WIMP can be a neutralino, 
which is the dark-matter candidate of the minimally supersymmetric (SUSY) SM. 
The neutralino-nucleon cross section is particularly sensitive to the strange scalar 
density, namely, the value of 
$y=2 \langle N | \bar s s | N \rangle/
\langle N | \bar u u + \bar d d | N \rangle$~\cite{Ellis}; its value impacts where 
 the allowed region of SUSY cross sections sit with respect to the
exclusion curves. Earlier studies relate $y$ to the $\pi N$ sigma term $\Sigma_{\pi N}$ via 
$y=1- \sigma_0/\Sigma_{\pi N}$ for fixed 
$\sigma_0 \equiv ((m_u+m_d)/2)
\langle N | \bar u u + \bar d d - 2 \bar s s | N \rangle$~\cite{Ellis,Giedt}. 
However, as Young has discussed at this meeting, $y$ can be computed directly 
using lattice QCD techniques and need not be predicated by the value of 
$\Sigma_{\pi N}$~\cite{Young,Giedt}. The outcome tends to make the spin-independent
WIMP-nucleon cross section smaller than that usually assumed~\cite{Giedt}, so that it would
tend to diminish the new physics reach of 
a particular WIMP direct detection experiment, note Ref.~\cite{dmtools} for comparisons
of experimental exclusion plots with theory. 
It is worth noting that an exclusion limit itself contains astrophysical 
assumptions~\cite{golwala}, 
so that the negative implications of flavor violation in this context for new physics
searches need not be definitive. 

\subsection{Flavor and CP Violation}

In studies of weak decays, flavor physics 
can impact or has impacted the discovery of CP-violating effects, 
the interpretation of CP-violating observables, and 
the pattern of new physics itself. 

My own work in this subject began with a visit to Adelaide in 1996, 
and I am here now in appreciation of Tony's gift of 
opportunity --- and to say a bit about how it has all 
turned out. In 1996 Tony and I had a common interest in isospin-breaking in the NN force;
in the language of meson-exchange currents, $\rho^0-\omega$ mixing 
mediates important 
isospin-breaking effects. We became intrigued by the notion of the role of 
$\rho^0-\omega$ mixing in hadronic B-meson decay~\cite{Enomoto:1996cv}; at the time
the asymmetric B-factories were not yet in operation. 
Our paper~\cite{Gardner:1997yx}, with Heath B. O'Connell, entitled 
``$\rho$-$\omega$ Mixing and Direct $CP$ Violation in 
Hadronic B Decays,'' proposed that 
the sign of the large CP-violating asymmetry predicted to exist 
in $B^\pm \to \rho^{\pm}\rho^0(\omega)$ decay could be used to fix 
the sign of $\sin\alpha$, where $\alpha=\hbox{arg}[-(V_{\rm td} V_{\rm tb}^\ast)
(V_{\rm ud} V_{\rm ub}^\ast)]$. 
This is possible because $\rho^0-\omega$ mixing 
in $B^\pm \to \rho^\pm \rho^0$ decay allows direct CP violation 
to occur, and information on the
needed mixing matrix element can be gleaned from 
$e^+e^-\to \pi^+ \pi^-$ data~\cite{Gardner:1997yx,Gardner:1997ie}. 
Let us now turn to the context
in which this work sits, as well as to a broader discussion of the role of flavor. 
\begin{figure}
  \includegraphics[height=.4\textheight]{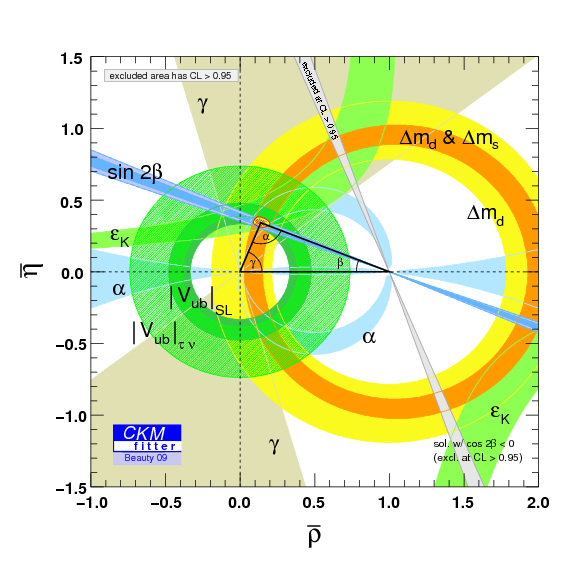}
  \caption{Empirical constraints on the apex of the unitarity triangle, $(\bar \rho, \bar \eta)$,
as of Beauty 2009 from the CKMfitter group~\cite{ckmfitter}.}
\label{current}
\end{figure}

In the SM all CP-violating, 
weak-interaction phenomena derive from a single phase in 
the Cabibbo-Kobayashi-Maskawa (CKM) matrix~\cite{CKM}. 
To test the mechanism of CP violation in the SM we 
must thus test the relationships it entails. 
CP-violating effects 
first appear at 
one-loop order in the weak interaction because all three generations 
must participate. However, 
intergenerational mixing is suppressed by factors of $|V_{\rm us}|\equiv \lambda 
\sim 0.2$ in the SM, so that 
CP-violating effects can be difficult to detect experimentally. 
From this perspective B-meson decays offer 
a particularly auspicious way to study CP violation because 
the requisite quark mixing effects are all of ${\cal O}(\lambda^3)$, 
and CP violation can appear without suppression from intergenerational mixing. 
The CKM unitarity test probed 
by $b$-quark decay reflects this as well; thus enters ``the'' unitarity triangle formed from the
constraint $V_{\rm ub}^\ast V_{\rm ud}^{} + V_{\rm cb}^\ast V_{\rm cd}^{} + 
V_{\rm tb}^\ast V_{\rm td}^{} =0$. The current empirical status of the
determination of the apex of the unitarity triangle is shown in 
Fig.~\ref{current}. The height of the apex is of ${\cal O}(1)$; thus CP-violating
effects can be large in the SM, though they appear in a rather special way. 

Different CP-violating effects exist in the B meson system~\cite{Bigi:1981qs,Nir:1992wi}. 
CP violation in the B system can appear in $B^0\bar B^0$
mixing, though this has not yet been observed. If we consider $B^0$ decay to 
a CP self-conjugate final state $f_{\rm CP}$, it can also appear as 
the interference between $B^0\bar B^0$ mixing and direct decay. 
A meson which is ``tagged'' as a $B^0$ meson at proper time $\tau=0$ 
has a finite probability of being a $\bar B^0$ meson at proper time $\tau$, 
so that both $B^0\to f_{\rm CP}$ and $B^0\to \bar B^0 \to f_{\rm CP}$ 
can occur. If the decay $B^0, \bar B^0\to f_{\rm CP}$ is controlled by
a single weak phase, then all strong-interaction effects cancel in 
the time-dependent, CP-violating asymmetry which yields 
$\sin 2\phi_B$ where $\phi_B$, the phase of $B^0\bar B^0$ mixing, is $\beta$ in 
the SM. 
Finally, CP violation can also appear in direct decay; this can be probed 
through a non-zero value of the partial-rate asymmetry, i.e., via 
$|A(B\to h_1 h_2)|^2 - |A(\bar B\to \bar h_1 \bar h_2)|^2 \ne 0$. 

Figure~\ref{current} shows that the CKM mechanism of CP violation
works very well; this sets important constraints on the energy scale of possible
new physics~\cite{nir2}. We see, moreover, that 
the determinations of $\beta$ and $\alpha$ are both discretely ambiguous, so that 
resolving a discrete ambiguity 
can potentially be an efficient means of identifying new physics. 
Interestingly, our suggestion~\cite{Gardner:1997yx} of 
resolving the sign of $\sin \alpha$ is 
still relevant!

To illustrate the role of flavor in the study of CP violation, 
we will consider how flavor informed the discovery of direct CP violation in the B system 
as well as how it can be used to make sharper tests of the SM of CP violation. 

\paragraph{On direct CP violation}
In the B-meson system direct CP violation was first observed in 2004 via
the rate asymmetry in $B\to K^+ \pi^-$~\cite{Aubert:2004qm}. 
To leading order in $G_F$, the sign of the $K^\pm$ ``tags'' the
flavor of the decaying $B^0$ ($\bar B^0$) meson, so that one loses none of the
data set to the determination of the flavor of the decaying B-meson. 
The measured rate asymmetry is~\cite{Aubert:2004qm} 
${\cal A}_{K\pi} 
\equiv {(n_{K^- \pi^+} - n_{K^+ \pi^-})}/{(n_{K^- \pi^+} + n_{K^+ \pi^-})} 
= -0.133 \pm 0.030 (\hbox{stat}) + 0.009 (\hbox{sys})\,$,
in good agreement with Ref.~\cite{Chao:2004mn}, 
where $n_{K^- \pi^+}$ is the measured yield for the $K^- \pi^+$ final state. 
The size of the asymmetry is uncomfortably large for QCD factorization~\cite{Beneke:2003zv}, 
but the empirical asymmetry can be confronted successfully without 
 physics beyond the SM~\cite{Buras:2004th,Brodsky:2001yt}. 
The observation of CP violation in the D meson system, however, 
would signal new physics. Analogous ``untagged'' studies of direct CP
violation in D decay are possible, such as in $D\to K_S \pi^+ \pi^-$, e.g., or to 
any final state of definite CP in particle content. 
The breaking of the mirror symmetry of the untagged Dalitz plot realized in
terms of the $(K_S \pi^\pm)$ invariant masses 
would signal
direct CP violation~\cite{Gardner:2002bb,Gardner:2003su}. 

\paragraph{On the interpretation of CP-violating observables}
Flavor-based assumptions can be used to eliminate some sensitivity
to Standard Model physics but can induce other uncertainties. 
Let us first consider the measurement of $\beta$ with $b \to s$ penguins, to 
end of determining whether the value of $\sin(2\beta)$ is universal 
as in the SM~\cite{Grossman:1996ke}. 
In the case of the ``golden'' mode $B \to J/\psi K_S$, the direct decay 
is characterized by a single weak phase to an excellent approximation, 
so that the time-dependent CP asymmetry measures 
$\sin(2\beta)$.
For other modes, computation, or estimation, of the SM-induced shift
from $\sin(2\beta)$ is crucial.
In $B\to \phi K_S$~\cite{Grossman:1996ke,rupak} 
and $B\to f_0(980) K_S$~\cite{Dutta:2008xw} decays the corrections are small. 
This emerges in the former case 
because the $\phi$ is very nearly an ideally mixed ($s \bar s$) state, though 
the corrections to this limit have an appreciable effect~\cite{rupak}, 
and in the latter case because the vacuum-to-scalar matrix element of the 
strange vector current, which drives the ``wrong-phase'' tree 
contribution, vanishes by charge-conjugation symmetry~\cite{Dutta:2008xw,rupak}. 

We can also consider the measurement of $\alpha$ --- or $\gamma$, given $\beta$ ---
with the $b\to u$ amplitude through the $B\to \pi\pi$, or generally $B\to n\pi$, decay modes. 
Here an assumption of isospin symmetry is essential~\cite{Gronau:1990ka,Snyder:1993mx}.
Isospin-breaking effects are particularly important in 
$B\to \pi\pi$. In this case transition amplitudes can emerge which did not appear
in the isospin symmetric limit~\cite{Gardner:1998gz}. 
It is possible to assess all leading, strong isospin-breaking 
effects in the $B\to n\pi$ 
modes~\cite{Gardner:1998gz,Gardner:2001gc,Gronau:2005pq,Gardner:2005pq}. 
However, the current limits on $\alpha$ are driven by features of 
the $B\to\rho\rho$ data which are insensitive to isospin breaking~\cite{ckmfitter}. 

And what of the flavor of possible new physics? 
At current levels of precision, the CKM mechanism dominates flavor 
violation as well~\cite{ckmfitter,nir2}.

\section{Summary}

The pattern of CP violation in Nature
can be described by a single
parameter in the quark-mixing (CKM) matrix 
to an accuracy of some 20\%~\cite{RMP}. 
No new sources of CP 
violation beyond the SM have as yet been established. 
One can find terrestrial evidence for physics beyond the SM in low-energy
experiments by observing  
either  processes which are highly forbidden in the SM, such as 
CP violation in the D system or a permanent electric dipole moment, 
and/or by decided failures of unitarity triangle tests. The prospect
of the direct detection of dark matter also tantalizes. 
Flavor physics has played and will continue to play 
a crucial role in all of these developments.


\begin{theacknowledgments}
I would like to thank my collaborators for our work together and 
the organizers for the opportunity to participate in a workshop
on the occasion of Tony Thomas' 60$^{\rm th}$ birthday. 
I acknowledge partial support from the U.S. Department of Energy under 
contract DE-FG02-96ER30989. 
\end{theacknowledgments}



\bibliographystyle{aipproc}   

\begin{thebibliography}{9}

\bibitem{Komatsu:2010fb}
  E.~Komatsu {\it et al.},
  arXiv:1001.4538 [astro-ph.CO].

\bibitem{Feng:2010gw}
 J.~L.~Feng,
  arXiv:1003.0904 [astro-ph.CO] and references therein. 

\bibitem{Kaplan:1991ah}
  D.~B.~Kaplan,
  \emph{Phys.\ Rev.\ Lett.}  \textbf{68}, 741 (1992).

\bibitem{Londergan} J.~T. Londergan, these proceedings. 

\bibitem{Cloet} I.~C. Clo\"et, W. Bentz, and A.~W. Thomas, these proceedings. 

\bibitem{nutev} 
  G.~P.~Zeller {\it et al.}  [NuTeV Collaboration],
  \emph{Phys.\ Rev.\ Lett.}  \textbf{88}, 091802 (2002)
  [\emph{Erratum-ibid.}  \textbf{90}, 239902 (2003)]. 

\bibitem{Young} R.~D. Young, these proceedings. 

\bibitem{gaitskell} R.~J. Gaitskell, \emph{Ann.\ Rev.\ Nucl.\ Part.\ Sci.} \textbf{54},
315 (2004). 

\bibitem{Ellis} 
  J.~R.~Ellis, K.~A.~Olive, and C.~Savage,
  \emph{Phys.\ Rev.\  D} \textbf{77}, 065026 (2008). 

\bibitem{Giedt} 
  J.~Giedt, A.~W.~Thomas, and R.~D.~Young,
 \emph{Phys.\ Rev.\ Lett.}  \textbf{103}, 201802 (2009).

\bibitem{dmtools} R. Gaitskell, V. Mandic, and J. Filippini, for
updated results and plots see:  \url{http://dmtools.brown.edu}. 

\bibitem{golwala} S.~R. Golwala, Ph.\ D.\ thesis, 
\url{http://cosmology.berkeley.edu/preprints/cdms/golwalathesis/}.

\bibitem{ckmfitter} 
CKMfitter Group (J. Charles et al.),
\emph{Eur. Phys. J.} \textbf{C41}, 1-131 (2005), 
[hep-ph/0406184],
updated results and plots available at: \url{http://ckmfitter.in2p3.fr}.

\bibitem{Enomoto:1996cv}
  R.~Enomoto and M.~Tanabashi,
 \emph{Phys.\ Lett.\  B} \textbf{386}, 413 (1996). 

\bibitem{Gardner:1997yx}
  S.~Gardner, H.~B.~O'Connell, and A.~W.~Thomas,
 \emph{Phys.\ Rev.\ Lett.}  \textbf{80}, 1834 (1998). 

\bibitem{Gardner:1997ie}
  S.~Gardner and H.~B.~O'Connell,
  \emph{Phys.\ Rev.\  D} \textbf{57}, 2716 (1998)
  [\emph{Erratum-ibid.\  D} \textbf{62}, 019903 (2000)].

\bibitem{CKM} N. Cabibbo, \emph{Phys. Rev. Lett.} \textbf{10}, 531 (1963); 
M. Kobayashi and T. Maskawa, \emph{Prog. Theor. Phys.} \textbf{49}, 652 (1973). 

\bibitem{Bigi:1981qs}
  I.~I.~Y.~Bigi and A.~I.~Sanda,
  \emph{Nucl.\ Phys.\  B} \textbf{193}, 85 (1981).

\bibitem{Nir:1992wi}
  Y.~Nir and H.~R.~Quinn,
  \emph{Ann.\ Rev.\ Nucl.\ Part.\ Sci.}  \textbf{42}, 211 (1992) and references therein.

\bibitem{nir2} For a review, see 
  G.~Isidori, Y.~Nir, and G.~Perez,
  arXiv:1002.0900 [hep-ph].
 
\bibitem{Aubert:2004qm}
  B.~Aubert {\it et al.}  [BaBar Collaboration],
  \emph{Phys.\ Rev.\ Lett.} \textbf{93}, 131801 (2004).

\bibitem{Chao:2004mn}
  Y.~Chao {\it et al.}  [Belle Collaboration],
   \emph{Phys.\ Rev.\ Lett.} \textbf{93}, 191802 (2004). 

\bibitem{Beneke:2003zv}
  M.~Beneke and M.~Neubert,
 \emph{Nucl.\ Phys.\  B} \textbf{675}, 333 (2003). 

\bibitem{Buras:2004th}
  A.~J.~Buras, R.~Fleischer, S.~Recksiegel, and F.~Schwab,
 \emph{Acta Phys.\ Polon.\  B} \textbf{36}, 2015 (2005). 

\bibitem{Brodsky:2001yt}
  S.~J.~Brodsky and S.~Gardner,
 \emph{Phys.\ Rev.\  D} \textbf{65}, 054016 (2002). 

\bibitem{Gardner:2002bb}
  S.~Gardner,
 \emph{Phys.\ Lett.\  B} \textbf{553}, 261 (2003). 

\bibitem{Gardner:2003su}
  S.~Gardner and J.~Tandean,
 \emph{Phys.\ Rev.\  D} \textbf{69}, 034011 (2004). 

\bibitem{Grossman:1996ke}
  Y.~Grossman and M.~P.~Worah,
  \emph{Phys.\ Lett.\  B} \textbf{395}, 241 (1997).

\bibitem{Grossman:1997gr}
  Y.~Grossman, G.~Isidori, and M.~P.~Worah,
  \emph{Phys.\ Rev.\  D} \textbf{58}, 057504 (1998).

\bibitem{rupak} R. Dutta and S. Gardner, in preparation. 

\bibitem{Dutta:2008xw}
  R.~Dutta and S.~Gardner,
  \emph{Phys.\ Rev.\  D} \textbf{78}, 034021 (2008).

\bibitem{Gronau:1990ka}
  M.~Gronau and D.~London,
  \emph{Phys.\ Rev.\ Lett.} \textbf{65}, 3381 (1990).

\bibitem{Snyder:1993mx}
  A.~E.~Snyder and H.~R.~Quinn,
  \emph{Phys.\ Rev.\  D} \textbf{48}, 2139 (1993).

\bibitem{Gardner:1998gz}
  S.~Gardner,
  \emph{Phys.\ Rev.\  D} \textbf{59}, 077502 (1999).

\bibitem{Gardner:2001gc}
  S.~Gardner and U.-G.~Mei{\ss}ner,
  \emph{Phys.\ Rev.\  D} \textbf{65}, 094004 (2002).

\bibitem{Gronau:2005pq}
  M.~Gronau and J.~Zupan,
  \emph{Phys.\ Rev.\  D} \textbf{71}, 074017 (2005).

\bibitem{Gardner:2005pq}
  S.~Gardner,
 \emph{Phys.\ Rev.\  D} \textbf{72}, 034015 (2005).

\bibitem{RMP}
T.~E.~Browder, T.~Gershon, D.~Pirjol, A.~Soni, and J.~Zupan,
\emph{Rev.\ Mod.\ Phys.} \textbf{81}, 1887 (2009).


\end{thebibliography}


\end{document}